\documentclass[a4paper]{article}
\newcommand{\be} {\begin{equation}}
\newcommand{\ee} {\end{equation}}
\newcommand{\bdm} {\begin{displaymath}}
\newcommand{\edm} {\end{displaymath}}
\newcommand{\bc} {\begin{center}}
\newcommand{\ec} {\end{center}}
\newcommand{\beqa} {\begin{eqnarray}}
\newcommand{\eeqa} {\end{eqnarray}}

\newcommand{\bfig} {\begin{figure}}
\newcommand{\efig} {\end{figure}}
\newcommand{\btab} {\begin{tabular}}
\newcommand{\etab} {\end{tabular}}

\usepackage{epsfig}
\begin{document}
\bc
{\bf\Large Evidence for Quark Spin-flip in Pomeron Exchange}
\ec
\bigskip
\bc
{\large A Donnachie}\\
{\large Department of Physics and Astronomy, University of Manchester}\\
{\large Manchester M13 9PL, England}
\ec
\medskip

\begin{abstract}
\noindent Spin-parity analyses of the $\omega\pi$ system in the reaction 
$\gamma p \to (\omega\pi) p$ for photon laboratory energies from 20 to 70 
GeV have shown that production of the $J^P=1^+$ $b_1(1235)$ meson dominates, 
with a $J^P=1^-$ background at the level of $20\%$. Using vector-meson 
dominance arguments, this background is shown to be consistent with the 
data on $e^+e^- \to \omega\pi$. The energy dependence of the data imply 
that the mechanism is a combination of Reggeon and Pomeron exchange. 
Assuming that the latter is relevant only for the $J^P = 1^-$ component 
and extrapolating to $W=200$ GeV, it is argued that this accounts for most 
of the preliminary $\omega\pi$ signal observed by the H1 Collaboration in 
the same reaction. A residual peak can be ascribed to the $b_1(1235)$, 
which requires a quark spin-flip from Pomeron exchange. Precisely the 
same mechanism occurs in the reaction $\pi p \to a_1(1260) p$.
\end{abstract}  

\bigskip

{\large
Preliminary data from the H1 Collaboration \cite{H102} on the reaction 
$\gamma p \to (\omega\pi^0) X$ at $\langle W \rangle = 200$ GeV and 
$\langle W \rangle = 210$ GeV was provisionally interpreted as diffractive 
$b_1(1235)$ production. After subtraction of the non-resonant background 
predicted by Pythia, the cross section for $\gamma p \to b_1(1235) X$ is
\be
\sigma(\gamma p \to b_1(1235) X) = 790 \pm 200({\rm stat})\pm 
200({\rm syst})~~{\rm nb}
\ee
At first sight it is unlikely that the $b_1(1235)$ can be produced by 
Pomeron exchange, which this interpretation requires. The transition 
$\gamma \to b_1(1235)$ does not satisfy the Gribov-Morrison rule 
\cite{Gri67,Mor68} which relates the change in spin $\Delta J$ to the 
change in parity between the incident particle and the outgoing resonance 
by $P_{\rm out}=(-1)^{\Delta_J}P_{\rm in}$. Further it is well-known 
experimentally that pomeron exchange conserves helicity to a good 
approximation, so that helicity-flip amplitudes are small. This is 
in agreement with the phenomenological $\gamma_\mu$ coupling of the 
pomeron to quarks \cite{LP71}. The $q\bar{q}$ pair from a photon are in 
a spin-triplet state, as exemplified by vector-meson dominance, but the 
quarks in the $b_1(1235)$ meson are in a spin-singlet state so quark 
helicity flip is required for the $\gamma \to b_1(1235)$ transition. 
There is also experimental evidence, at lower energy, that the reaction 
$\gamma p \to (\omega\pi^0) p$ is not dominated by pomeron exchange. 

The Omega Photon Collaboration \cite{CERN1} at CERN performed a spin-parity 
analysis of the $\omega\pi^0$ enhancement photoproduced in the energy range 
20 to 70 GeV, with $\langle W \rangle = 8.6$ GeV. They concluded that the 
enhancement is consistent with predominant $b_1(1235)$ production, with 
$\sim 20\%$ $J^P = 1^-$ background. This conclusion was confirmed by a SLAC 
experiment \cite{SLAC} at an energy of 20 GeV, $W = 6.2$ GeV, with a 
polarised beam. It should be noted that a spin-parity analysis of the H1 
data cannot be performed because of limited acceptance. It was possible to 
measure the energy dependence of the reaction in the CERN experiment, with 
the result
\be
\sigma(E_\gamma) = \sigma(39)\Big(\frac{39}{E_\gamma}\Big)^\alpha,~~~~20 
\le E_\gamma \le 70~~{\rm GeV}
\label{sig1}
\ee
with
\be
\sigma(39) = 0.86 \pm 0.27 \mu{\rm b},~~~~\alpha = 0.6 \pm 0.2
\label{sig2}
\ee
Such an energy dependence is not consistent with dominance of pomeron 
exchange, which would require an increasing cross section, nor is it 
consistent with pure Regge exchange, which would require a somewhat faster 
decrease with increasing energy. A natural interpretation is that the 
observed energy dependence arises from a combination of pomeron and 
reggeon exchange. As a simple first approximation, consider the cross
section to be given by non-interfering reggeon and pomeron exchanges, the 
former relating primarily to $b_1(1235)$ production and the latter relating 
entirely to the production of the $J^P = 1^-$ state. The energy dependence 
of the cross section (\ref{sig1}) can be well reproduced by 
\be
\sigma(s) = As^{2\epsilon}+Bs^{-2\eta}
\label{sig3}
\ee
where $\epsilon$ and $\eta$ have the standard values \cite{DL} 0.08 and 0.4525
respectively and
\be
A = 0.107~\mu{\rm b},~~~~B = 29.15~\mu{\rm b}
\label{sig4}
\ee
At $E_\gamma = 39$ GeV the pomeron contribution to the cross section is 
25$\%$, in good agreement with what is observed for the $J^P = 1^-$ component 
in the data. Extrapolating the pomeron part of (\ref{sig3}) to HERA energies 
gives a cross section of 584 nb. As the HERA data include diffraction 
dissociation of the nucleon, the result extrapolated from the fit to the 
CERN data should be increased by a factor of about 1.25 giving 730 nb, 
compatible with the cross section observed. The reggeon part of the cross 
section is negligible at this energy. 

\bfig[t]
\bc
\begin{minipage}{65mm}
\epsfxsize65mm
\epsffile{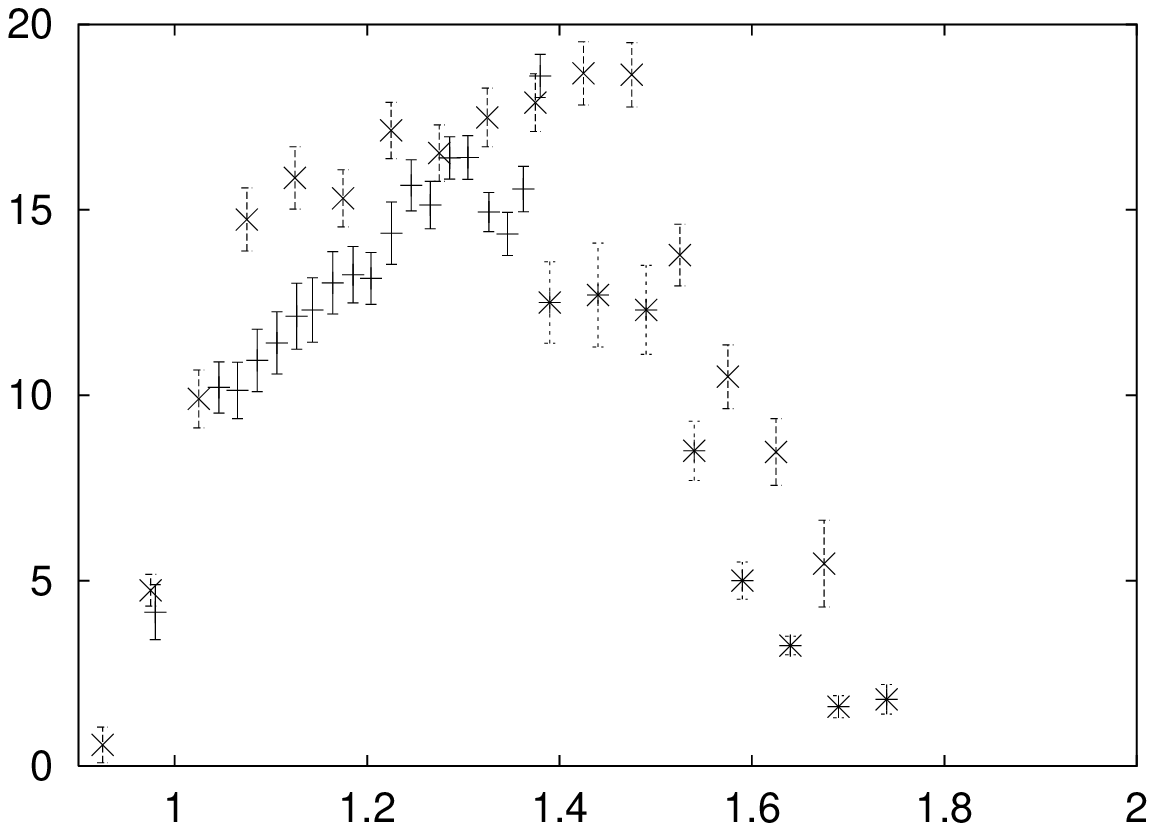}
\begin{picture}(0,0)
\setlength{\unitlength}{1mm}
\put(50,40){\small{(a)}}
\put(-5,40){\small{$\sigma$ (nb)}}
\put(30,-2){\small{$m$ (GeV)}}
\end{picture}
\end{minipage}
\hfill
\begin{minipage}{65mm}
\epsfxsize65mm
\epsffile{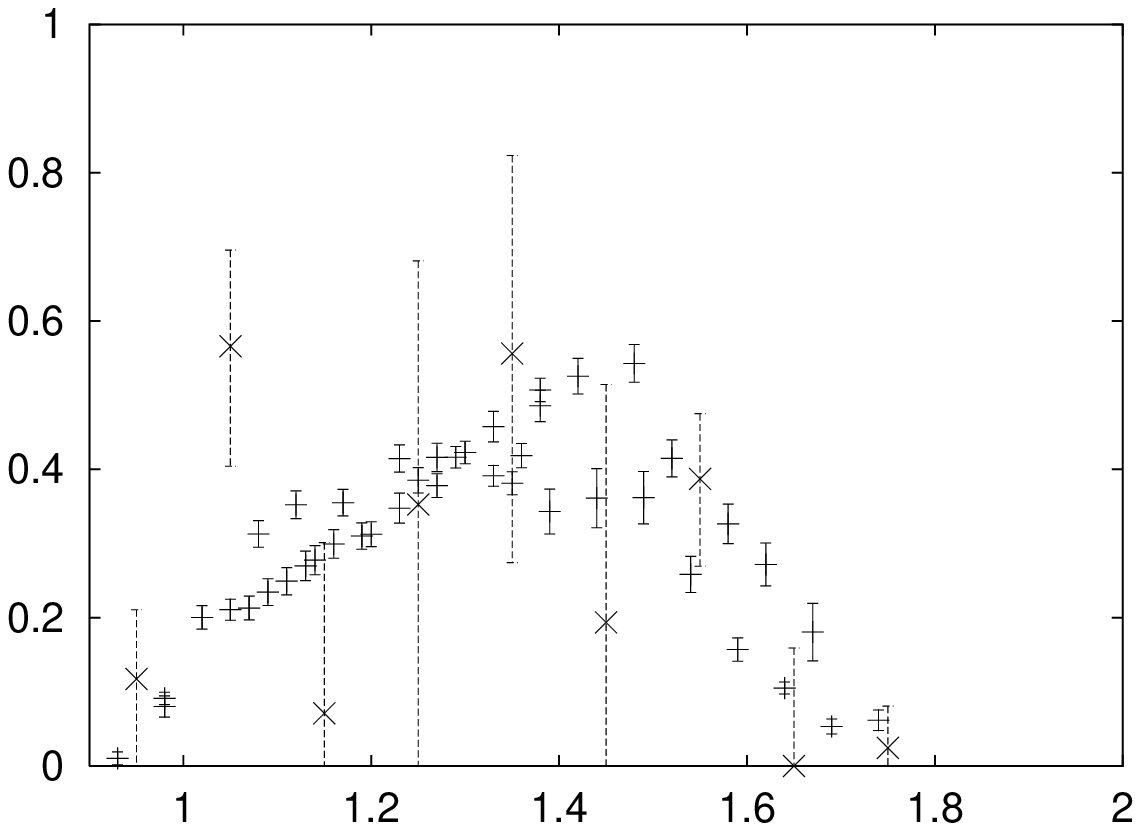}
\begin{picture}(0,0)
\setlength{\unitlength}{1mm}
\put(50,40){\small{(b)}}
\put(-5,40){\small{$\frac{d\sigma}{dm}$}}
\put(-8,35){\small{($\mu$b/GeV)}}
\put(30,-2){\small{$m$ (GeV)}}
\end{picture}
\end{minipage}
\caption{(a) The cross section for $e^+e^- \to \omega\pi$. The data are from 
Novosibirsk \cite{novo} (horizontal bars), CLEO \cite{cleo} (crosses) and
the DM2 Collaboration \cite{dm2} (stars)  (b) The $J^P = 1^-$ component of
the $\omega\pi$ mass distribution in the reaction $\gamma p \to \omega\pi p$ 
at $\sqrt{s}=8.5$ GeV. The data are from the Omega Photon Collaboration 
\cite{CERN1} (crosses) and from the application of vector meson dominance 
to the data in (a) (horizontal bars).} 
\ec
\efig

What is the origin of this $J^P = 1^-$ component? An estimate can be made 
using simple vector meson dominance arguments. For a vector final state $V$, 
the cross section for $\gamma p \to V p$ is related to that for $e^+e^- \to V$ 
by \cite{SS72}
\be
\frac{d^2\sigma_{\gamma p \to V p}(s,m^2)}{dt~dm^2} = \frac{\sigma_{e+e- \to V}
(m^2)}{4\pi^2\alpha}\frac{d\sigma_{V p \to V p}(s,m^2)}{dt}
\label{sig5}
\ee
Using the optical theorem to relate the amplitude at $t=0$ to the total cross 
section for $V p$ scattering and integrating over $t$ gives
\be
\frac{d\sigma_{\gamma p \to V p}(s,m^2)}{dm} = \frac{m\sigma_{e+e- \to V}(m^2)}
{32 \pi^3 \alpha b}(\sigma^{Tot}_{V p \to V p}(s))^2
\label{sig6}
\ee
where $b \approx 5$ GeV$^{-2}$ is the slope of the near-forward differential 
cross section.

The cross section for $\gamma p \to \pi^+\pi^-\pi^+\pi^- p$ over the same 
energy and four-pion mass ranges as the $\omega\pi$ photoproduction data has 
been compared with the data on $e^+e^- \to \pi^+\pi^-\pi^+\pi^-$ by the Omega 
Photon Collaboration \cite{CERN2}. This gave the result 
\be
\sigma^{Tot}_{V p \to V p} = 16.7 \pm 3.4~{\rm mb.}
\label{sig7}
\ee
Three models were considered in the spin-parity analysis \cite{CERN1} of the 
$\gamma p \to \pi^+\pi^-\pi^+\pi^- p$ data: $\omega\pi$ states with $J^P = 1^+,
 1^-, 0^-$, $J^P = 1^+, 1^-$ and $J^P = 1^+, 1^-$ with the $1^-$ constrained 
to be $s$-channel helicity conserving. It is the third one that we use here.
 
The data \cite{novo,cleo,dm2} for $e^+e^- \to \omega\pi$ are shown in Fig.1a 
and the comparison with $d\sigma/dm$ in Fig.1b. The errors arising from 
(\ref{sig7}) have not been included. The normalisation in this comparison is 
absolute and shows that the model produces the same $J^P=1^-$ cross section 
as the reggeon plus pomeron fit, within the admittedly large errors.

\bfig[t]
\bc
\begin{minipage}{70mm}
\epsfxsize70mm
\epsffile{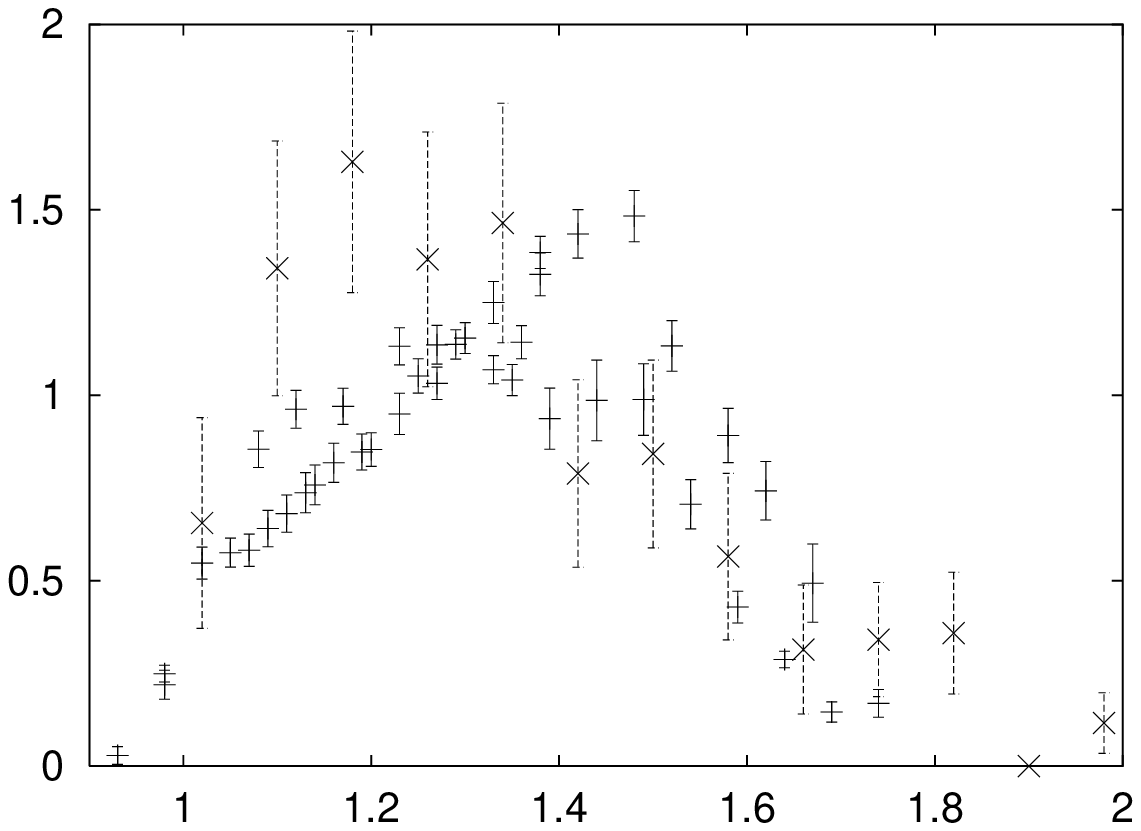}
\begin{picture}(0,0)
\setlength{\unitlength}{1mm}
\put(-5,40){\small{$\frac{d\sigma}{dm}$}}
\put(-9,35){\small{($\mu$b/GeV)}}
\put(30,-2){\small{$m$ (GeV)}}
\end{picture}
\end{minipage}
\caption{The $\omega\pi$ mass distribution in the reaction $\gamma p \to 
\omega\pi p$ at $\sqrt{s}=200$ GeV. The data (preliminary) are from the 
H1 Collaboration \cite{H102} (crosses) and from the application of vector 
meson dominance to the data in Fig.1a (horizontal bars).}
\ec
\efig

A similar comparison can be made with the HERA data and this is shown in 
Fig.2 after converting the preliminary H1 data from events/bin to $d\sigma/dm$ 
assuming 790 nb as the integrated cross section. At the upper end of the mass 
range the agreement is reasonably good, perhaps surprisingly so given the 
overall errors in the procedures we are using. However the overall shapes are 
not the same, with an apparent excess of H1 data at the lower mass end. This 
could be explained if there were some diffractive production of the 
$b_1(1235)$. As we said initially, diffractive production of the $b_1(1235)$ 
implies a spin-flip pomeron-exchange contribution. Is this reasonable?

\bfig[t]
\bc
\begin{minipage}{70mm}
\epsfxsize70mm
\epsffile{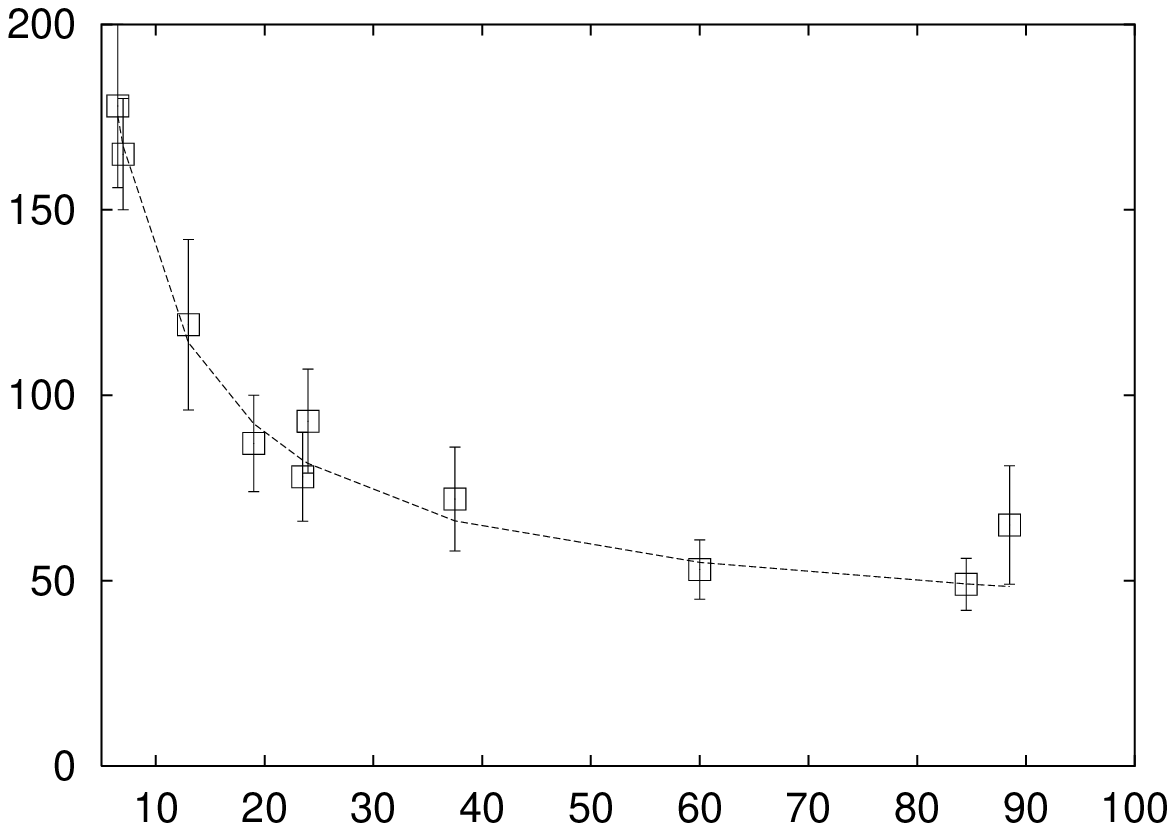}
\begin{picture}(0,0)
\setlength{\unitlength}{1mm}
\put(-5,42){\small{$\sigma$ ($\mu$b)}}
\put(30,-2){\small{$p_{\rm Lab}$ (GeV)}}
\end{picture}
\end{minipage}
\caption{The cross section for $\pi^- p \to a_1(1260) p$. The data are from the
ACCMOR Collaboration \cite{ACCMOR} and the curve is the fit using the 
parametrisation (\ref{a1-2})}
\ec
\efig

An analogous reaction is $\pi^- p \to a_1(1260) p$. Although this does satisfy 
the Gribov-Morrison rule, like the reaction $\gamma p \to b_1(1235)$ it 
requires spin-flip at the quark level. The data \cite{ACCMOR} are shown in 
Fig.3 and the reaction is clearly not pomeron dominated. Fitting with a single 
effective power,
\be
\frac{d\sigma}{dm} = As^\alpha
\label{a1-1}
\ee
gives $\alpha = 0.52$. This is very close to the value found for $\gamma p \to 
b_1(1235) p$ and it is natural to make the same interpretation, namely that the
reaction mechanism is a combination of reggeon and pomeron exchange. However in
this case, as we have only one final state, we must allow for interference. The
curve in Fig.3 is given by
\be
\sigma = As^{2\epsilon}+Bs^{\epsilon-\eta}+Cs^{-2\eta}
\label{a1-2}
\ee
with $A=7.87$ $\mu$b, $B = 98.6$ $\mu$b, $C = 1231$ $\mu$b. The values of 
$\epsilon$ and $\eta$ are the same as before. 

We can estimate the relative strength of the effective spin-flip coupling of 
the pomeron to the non-spin-flip coupling by comparing the fit (\ref{a1-2}) 
with a corresponding fit to the elastic $\pi p$ cross section. The result of 
such a fit for $p_{\rm Lab} \ge 4$ GeV gives $A_{\rm el}=1.025$ mb. Thus the 
contribution of pomeron exchange to the $\pi p$ elastic cross section 
(non-spin-flip) is a factor of 130 more than the contribution of the
pomeron exchange to the cross section for $\pi p \to a1(1260) p$ (spin flip). 
Note that the mechanism we are suggesting here is not the one responsible for 
the small violation of $s$-channel helicity conservation in $\gamma p \to 
\rho p$ \cite{ZEUS00,H100}. This small effect can be explained 
\cite{IK98,KNZ98} by a mechanism which conserves helicity at the quark level.

The difference in cross section beween the H1 data and the estimated $J^P=1^-$ 
contribution shown in Fig.2 is of the order of 0.1 - 0.2 $\mu$b with a large 
error due to the errors on both the photoproduction and $e^+e^-$ annihilation 
data and the simplicity of the model used to estimate the $J^P = 1^-$ 
contribution. The cross section for $\rho$ photoproduction at $W = 70$ GeV 
is given as $14.7 \pm 0.4 \pm 2.4$ $\mu$b by ZEUS \cite{ZEUS95} and as 
$13.6 \pm 0.8 \pm 2.4$ by H1 \cite{H195}. Extrapolating these to $W = 200$ 
GeV gives $20.5 \pm 3.4$ $\mu$b and $19 \pm 3.5$ $\mu$b respectively, where 
the statistical and systematic errors have been combined in quadrature. So 
the ratio of spin-flip to non-spin-flip pomeron exchange in $b_1(1235)$ 
photoproduction is of the same order of magnitude as in the case of the 
$a_1(1260)$. 

There are a few reactions in which this hypothesis can be checked. The ideal 
would be a new measurement of the energy dependence and full spin-parity 
analysis of $\omega\pi$ photoproduction. The photoproduction of the isoscalar 
counterpart of $\gamma p \to b_1(1235) p$, namely $\gamma p \to h_1(1170)$, 
$h_1 \to \rho\pi$, would be expected to occur at about $10\%$ of the 
$b_1(1235)$ photoproduction cross section, so would be of the order of 50 
to 100 nb at HERA energies. The hypothesis also provides a mechanism for 
diffractive photoproduction of the unconfirmed hidden-strangeness $h_1(1380)$ 
with a cross section at the level of $1\%$ of the $\phi$ photoproduction cross 
section, which is $0.96 \pm 0.19 ^{+0.21}_{-0.18}$ $\mu$b at $\langle W 
\rangle =70$ GeV \cite{ZEUSphi}. So we would expect about 10 nanobarns at 
this energy.

The suggestion that the pomeron may have a spin-flip coupling is not new and 
has been discussed extensively in the context of proton-proton scattering and 
diffractive hadron leptoproduction in a number of models. Probably the most 
relevant for the present context is a purely phenomenological approach 
\cite{mp02} to proton-proton scattering which concluded that there is a 
spin-flip pomeron amplitude with the same trajectory as the standard 
spin-non-flip pomeron. However this does require the inclusion of an 
arbitrary phase difference between the non-spin-flip and spin-flip components
of the pomeron so it is not a strict Regge-pole parametrisation. Spin-flip in 
diffractive reactions has also been discussed at the parton level, for example 
in proton-proton scattering at large $|t|$ \cite{gol02} and in vector-meson 
and $Q\bar Q$ production in deep inelastic scattering \cite{gol03,gkp03}. 
Although these are more applicable in the framework of perturbative QCD, 
they reach the same general conclusions.

}

{\large

}
\end{document}